\def\n{\noindent}

\def\be{\begin{equation}}
\def\ee{\end{equation}}

\def\x{{\underline{r}}}

\def\P{$P$}

\def\eq{\enskip =\enskip}
\def\pls{\enskip +\enskip}
\def\mns{\enskip -\enskip}

  \def\ket{\vert \vert  \{ \emptyset \} \rangle}
  \def\ket2{\vert \vert \otimes \{ R \} \rangle}

\def\.#1{\mathaccent 95#1}
\def\^#1{\mathaccent 94 #1}
\def\~#1{\mathaccent "7E #1}

\def\eq{\enskip =\enskip}
\def\pls{\enskip +\enskip}
\def\mns{\enskip -\enskip}

\def\un#1{\underline{#1}}

  \def\ket{\vert \vert  \{ \emptyset \} \rangle}
  \def\ket2{\vert \vert \otimes \{ R \} \rangle}

\def\r{\underline{r} }

\documentclass[12pt,epsfig]{iopart}
\usepackage{epsfig}
\begin{document}
\setcounter{page}{1}
\title{Study of a pair of coupled continuum equations modeling  surface growth}
\author{Ain-ul Huda\footnote{email : huda@bose.res.in}} 
\address{S.N. Bose National Centre for Basic Sciences, JD Block, Sector III, Salt Lake City,
Kolkata 700098, India}
\author{Omjyoti Dutta\footnote{email : omjyoti@rediffmail.com}}
\address{Department of Electronics and Telecommunications, Jadavpur University, Kolkata 700032, India}
\author{Abhijit Mookerjee\footnote{email : abhijit@bose.res.in}}
\address{S.N. Bose National Centre for Basic Sciences, JD Block, Sector III, Salt Lake City,
Kolkata 700098, India}

\begin{abstract}
In this communication we introduce a pair of coupled continuum equations to model
overlayer growth with evaporation-accretion due to thermal or mechanical agitations of the
substrate. We gain insight into the dynamics of growth via one-loop perturbative techniques.
This allows us to analyze our numerical data. We conclude that there is a crossover behaviour
from a roughening regime to a very long-time, large length scale smoothening regime.
\end{abstract}

\pacs{71.20, 71.20c}

\section{Introduction}
The dynamics of surface growth by atomic deposition have been the focus of interest
over recent years (\cite{intro1}-\cite{intro8}). 
 Several theoretical attempts (\cite{kn:st}) at the understanding of kinetic roughening
have been made, through discrete and continuum models, motivated by 
experiments.  Roughening is often an inevitable part of surface formation,  
so that an understanding of the surface morphologies has a crucial part to play in the many
 vital applications of this field.  However, not much attention has been paid  to the phenomenon
of smoothening by thermal effects like evaporation. The physical picture is clear :
the vibration of the substrate,  of thermal or mechanical origin may smoothen the overlayer 
surface by transferring weakly bonded atoms on  its bumps or mounds to available surface grooves. The
picture is similar, but certainly not equivalent, to smoothening of granular surfaces
by avalanches. This is in contrast to the {\sl surface diffusion} term in continuum
models, which arises because of the internal rearrangement of the {\sl bonded atoms} in
order to minimize the chemical potential. Avalanche smoothening  has been recently studied 
in some detail using coupled continuum equations by Biswas \etal (\cite{bmb}). 

The notion that we shall borrow and adapt  from the work of Biswas \etal is that the dynamics of atoms in
surface growth with evaporation-accretion is well described by the competition between the collective 
dynamics or relaxation of bonded atoms (in order to minimize the chemical potential) and the dynamics
of free atoms diffusing on  the surface. The bare surface of bonded, and therefore relatively 
immobile atoms, will be described by the local height $h(\un{r},t)$ above the substrate. Across
this surface the {\sl gas} of unbonded evaporated atoms  diffuse until they are {\sl captured} 
in an available groove. This {\sl gas} of atoms will be characterized by its density $\rho(\un{r},t)$ just above 
the bare surface. A similar model has been discussed by Sanyal \etal \cite{smm}. However, the 
transfer term in the equations was rather different from ours and the 
possibility of smoothening was not addressed in that work.

The usual practice for probing temporal or spatial roughness is to study the asymptotic behaviour of
 correlation functions like $\langle h(\un{r},t)h(\un{r'},t')\rangle$ in space or time via a single
Fourier transform. Only one of the variables : space or time, is integrated over in Fourier space and the
relevant scaling relations are used to determine the critical exponents that govern this behaviour. However,
as pointed out by Biswas \etal (\cite{bmb}), this leads to ambiguities for those problems where there may
be more than one scaling lengths. In such cases, the double Fourier transform provides a much deeper
insight. To clarify this point, let us introduce the salient features of such a study :

\n The connected two point self-correlation function of the local variable $A(\un{r},t)$, which can either be 
$h(\un{r},t)$ or $\rho(\un{r},t)$ is defined as :

\be 
S(\un{r}-\un{r}', t-t') \eq \langle A(\un{r},t) A(\un{r}',t')\rangle \mns \langle A(\un{r},t)\rangle\langle
 A(\un{r}',t')\rangle 
\ee

\n The scaling hypothesis implies that we have, in the absence of spatial anisotropy, 

\[ S(\un{R},0)\ \simeq\ \vert \un{R}\vert ^{2\alpha} \quad\quad \un{R}=\un{r}-\un{r}',
\quad R=\vert\un{R}\vert\quad \quad \mbox{and} \quad R \rightarrow\infty\]

\n for the saturated surface with $t>t_s$, where $t_s$ is the saturation time, and 

\[ S(0,\tau)\ \simeq\ \tau^{2\beta}\quad\quad \tau =\vert t-t'\vert \quad\quad \tau\rightarrow\infty\]

\n In general, 

\[ S(\un{R},\tau)\ \simeq\ R^{2\alpha}\ \Phi\left(\frac{\tau}{R^z}\right)\quad\quad \mbox{for both } R,\tau\rightarrow\infty\]

\n The scaling function $\Phi$ is universal, $\alpha$ is the roughness and $z=\alpha/\beta$  the dynamical exponent.
For the single Fourier transforms,

\[ S(\un{k},\tau=0)\ \simeq\ k^{-1-2\alpha} \quad\quad\mbox{for}\quad\quad k\rightarrow 0 \]

\n and

\be S(\un{R}=0,\omega)\ \simeq\ \omega^{-1-2\beta} \quad\quad\mbox{for}\quad\quad \omega\rightarrow 0 \label{sf}\ee

\n For the double Fourier transform, in case we assume {\sl strong scaling}. That is,  existence of 
single length and time scales, consequently a single dynamical exponent $z$.

\[ S(\un{k},\omega) \simeq \omega^{-1} k^{-1-2\alpha}\ \Psi\left(\frac{\omega}{k^z}\right) \quad\quad\mbox{for}\quad\quad
k,\omega\rightarrow 0\]

\n which gives,

\[ S(k,\omega=0)\ \simeq\ k^{-1-2\alpha-z}\quad\quad\mbox{for}\quad\quad k\rightarrow 0 \]
\n and
\be S(k=0,\omega)\ \simeq\ \omega^{-1-2\beta-1/z} \quad\quad\mbox{for}\quad\quad \omega\rightarrow 0 \label{df}\ee

It is important to examine  the single Fourier transforms in equations (\ref{sf}). For the calculation of $S(k,\tau=0)$ 
we need to take the {\sl saturated} surface after a long time, but for $S(R=0,\omega)$ we need to take the entire
growing surface but {\sl locally}. Thus $\alpha$ is related to the saturated and $\beta$ to the growing surface. However,
the double Fourier transform requires information of {\sl both} the growing and saturated surfaces, both
{\sl locally} and at large length scales. Biswas \etal (\cite{bmb})
correctly argue that, in case there are more than one length or time scales associated with the process, the double
Fourier transform should provide a much clearer picture.

\section{The Statistical Model}

\section{The statistical model for atomic deposition}

Atomic deposition has many features in common with granular deposition.
The added  feature is atomic binding. 
In the usual deposition geometry, a randomly fluctuating flux of atoms
is incident on a substrate. Atoms deposit on the surface of the substrate
and diffuse along it to  minimize the energy. A cloud of unbonded atoms
envelope this deposit and continuously exchange atoms with it through
evaporation and  re-deposition.

While non-equilibrium growth has been extensively studied by
coarse-grained classical stochastic equations (\cite{mln}), it is not
obvious {\it a priori} that the microscopic energetic constraints relevant to
atomic surfaces would automatically be satisfied by largely heuristic
classical terms.  In an earlier communication (\cite{smm}) we had presented electronic
energy calculations in support of our model of surface growth.

Among various physical processes which have been taken into account
in models of growing interfaces, {\it surface diffusion} has been
considered as the most important process involved. One
such model involves the linear fourth-order Mullins-Herring continuum
equation (\cite{kn:mh,kn:mul}) supported by the discrete model of Wolf and
Villain (WV) (\cite{kn:wv}) 

\begin{equation}
{\partial h(\un{r},t)} / {\partial t} = - D\ 
\nabla^{4}h(\un{r},t) + \eta(\un{r},t)
\end{equation}

\noindent where $h(\un{r},t)$ is the height of the interface from
some mean height $\langle h(\un{r},t)\rangle$ and $\eta(\un{r},t)$
represents Gaussian white noise as usual.  This equation yields a
large roughness exponent $\alpha$ = 1.5 in $d$=1.

In an earlier communication (\cite{smm}) we had presented a model
 to look at the effect of desorption or evaporation on 
relatively immobile atoms which
are bonded to the surface. A cloud of mobile
atoms above the surface  arise both from the impinging atomic beam
and from evaporation caused by atoms knocked out of the surface by
thermal or mechanical
disturbances. These are described by their local density
$\rho(\x,t)$.
We propose a new class of growth equations with an explicit coupling
between the profile of ``bonded'' atoms represented by the local height
of the surface $h({\bf x},t)$, and ``mobile'' atoms on the surface
represented by their
local density $\rho(\x,t)$. Our equations read: 

\begin{eqnarray}
{\partial h(\x,t) }/{\partial t} & = & - D_{h}
\nabla^{4} h(\x,t) - {\cal T} + \eta_{h}(\x,t)\nonumber \label{eq:one} \\
{\partial\rho(\x,t)}/{\partial t} & = & \phantom{-}  D_{\rho}\nabla^{2} \rho(\x,t) + {\cal T}
\label{eq:two}
\end{eqnarray}

\n where the transfer term $\cal{T}$ is given by :

\begin{equation} 
\fl {\cal T}  =  \nu  \left\vert \nabla^2 h(\r,t) \right\vert \left[1-\Theta(\nabla^2 h(\r,t))\right] \mns
\mu \rho(\r,t)  \left\vert \nabla^2 h(\r,t) \right\vert \ \Theta(\nabla^2 h(\r,t))
\label{eq:three}
\end{equation}

\n where $\Theta(x)$ is 1 for $x>0$ and 0 otherwise. 
 We describe in what follows the meaning of the above terms. 

 \begin{description}
\item[(i)] The fourth-order term in the equation(\ref{eq:one}) describes
surface diffusion of bonded atoms; this is the usual
WV (\cite{kn:wv}) term where $D_{h}$ represents a diffusivity. In the continuum
picture surface diffusion is a manifestation of the collective motion of the bonded
atoms leading to a shape rearrangement in order to minimize the chemical potential.
 The leading to  this term is the gradient of the local chemical potential, which is  
assumed to be proportional to the local curvature. This assumption was shown to be valid
provided we invoke the {\sl Locality Principle} of Heine (\cite{heine}).

\item[(ii)] The flowing atoms are neither bonded to one another nor to
atoms on the surface. The first term in equation (\ref{eq:two})
describes normal, as opposed to surface diffusion of these mobile atoms, where the
corresponding current is the gradient of the density.

\item[(iii)] The first term in the transfer term $\cal{T}$, equation (\ref{eq:three}), describes
spontaneous
generation of mobile atoms  on the surface through {\sl evaporation} or {\sl desorption}. This
could be due, for example, to thermal disturbances.
 We have assumed that it is easier to
eject atoms weakly bonded at spiky mounds on the surface. For thermal ejection, for example, 
$\nu$ is a measure of the substrate temperature. The Theta function ensures that it is easier to thermally
eject atoms bonded on negative curvatures, i.e. on bumps or mounds on the surface. We have
assumed that the rate of evaporation is proportional to the negative curvature on a mound.

\item[(iv)] The second term in $\cal{T}$, equation (\ref{eq:three}),  represents
 {\sl condensation}, whereby mobile atoms accumulate and accrete preferentially at points
of positive curvature, i.e. at the bottom of deep grooves where atomic coordination is large and
hence bonding is strong. This term is obviously also proportional to the local density of mobile atoms.
 
\item[(v)] Finally the last term  is a
Gaussian white noise characterized by its width $\Delta_{h}$

\[
   \langle\eta_{h}(\x,t)\eta_{h}(\x',t')\rangle  =
\Delta^{2}_{h}\ \delta(\x-\x')\delta(t-t')
\]

\end{description}

\n We assume that  growth occurs on a flat substrate;  this and  the absence of 
a preferred direction leads to the absence of anisotropy in space. In this model we have ignored
the effects of  Sch\"owbel barriers.

\begin{figure}[ht]
\begin{center}
\epsfxsize=4.5in \epsfysize=3.5in
\rotatebox{0}{\epsfbox{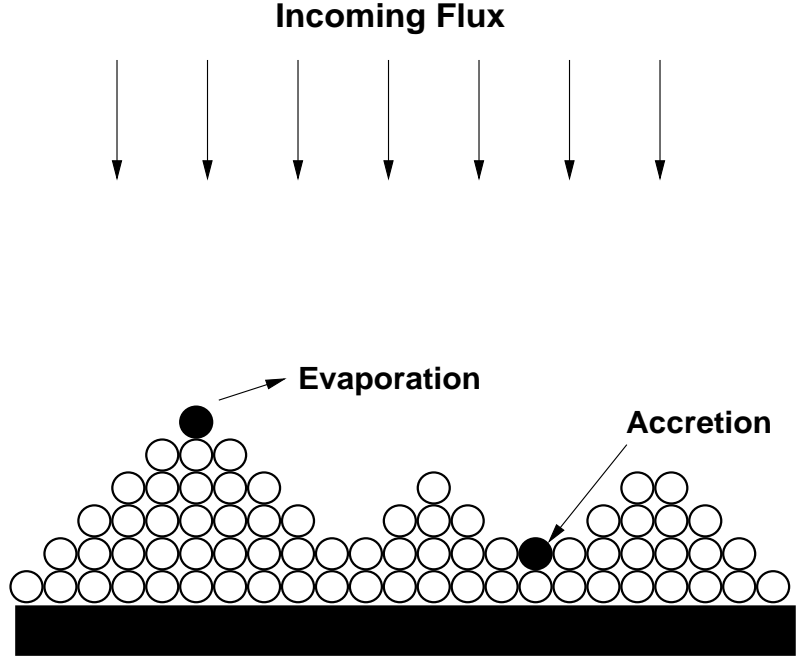}}
\caption{A pictorial depiction of the model}
\label{fig1}
\end{center}
\end{figure}

We can visualize the following sequence of processes: first, the
mobile atoms diffuse ($\nabla^2 \rho$) in the cloud above the surface.
This
is followed by the preferential conversion of these atoms into the
bonded species at points of high positive curvature ($\rho |\nabla^2 h|$) on the
surface such as mounds and grooves.  The term $\nu |\nabla^2 h|$ models the effect
of evaporation, leading to a dynamical
exchange, at regions of high negative curvature, between bonded and unbonded atoms.
However, the action of the $\nabla^4 h$ term is to stabilize the
formation of mounds and grooves, so ultimately the overwhelming effect is a
competition between roughening and smoothening  of the surface. 
Figure \ref{fig1} illustrates the effect of the terms in our model.

\section{Numerical and theoretical analysis}

To start with we shall analyze a model in (1+1) dimension. The substrate is one-dimensional, so that the position
of the overlayer on the substrate is located by a single position variable $x$. The coupled equations become :

\begin{eqnarray}
\fl \frac{\partial h(x,t)}{\partial t}  \eq  -\ D_h\ \frac{d^4 h(x,t)}{dx^4}\ -\ \nu\ \frac{d^2 h(x,t)}{dx^2}\ +\ \mu\ \rho(x,t)\frac{d^2 h(x,t)}{dx^2}\pls \eta_h(x,t)\nonumber \\    
\fl\phantom{x} \nonumber\\
\fl \frac{\partial \rho(x,t)}{\partial t}  \eq  \phantom{-}\ D_\rho\ \frac{d^2 \rho(x,t)}{dx^2}\ +\ \nu\ \frac{d^2 h(x,t)}{dx^2}\ -\ \mu\ \rho(x,t)\frac{d^2 h(x,t)}{dx^2} \label{req}     
\end{eqnarray}
\vskip 0.2cm

\n  The Heaviside step functions in equation (\ref{eq:three}) introduces a complexity in our
equations as far as analytic investigations are concerned. We shall follow the remarks of
Biswas \etal (\cite{bmb}) and invoke a suitable representation of the Heaviside function as
an infinite series. In that case, the equation (\ref{eq:three}) can be thought of as :

\begin{eqnarray} 
 {\cal T}  =  \nu  \frac{d^2 h(x,t)}{dx^2}  \mns
\mu\ \rho(x,t)  \frac{d^2 h(x,t)}{dx^2}\pls\ldots\nonumber\\
\ldots\pls \sum_{n=1}^{\infty} \nu_n \left( \frac{d^2 h(x,t)}{dx^2}\right)^n
\mns \sum_{n=1}^{\infty} \mu_n\ \rho(x,t) \left(\frac{d^2 h(x,t)}{dx^2}\right)^n
\end{eqnarray}

\begin{figure}[t]
\begin{center}
\epsfxsize=4in \epsfysize=4.5in
\rotatebox{270}{\epsfbox{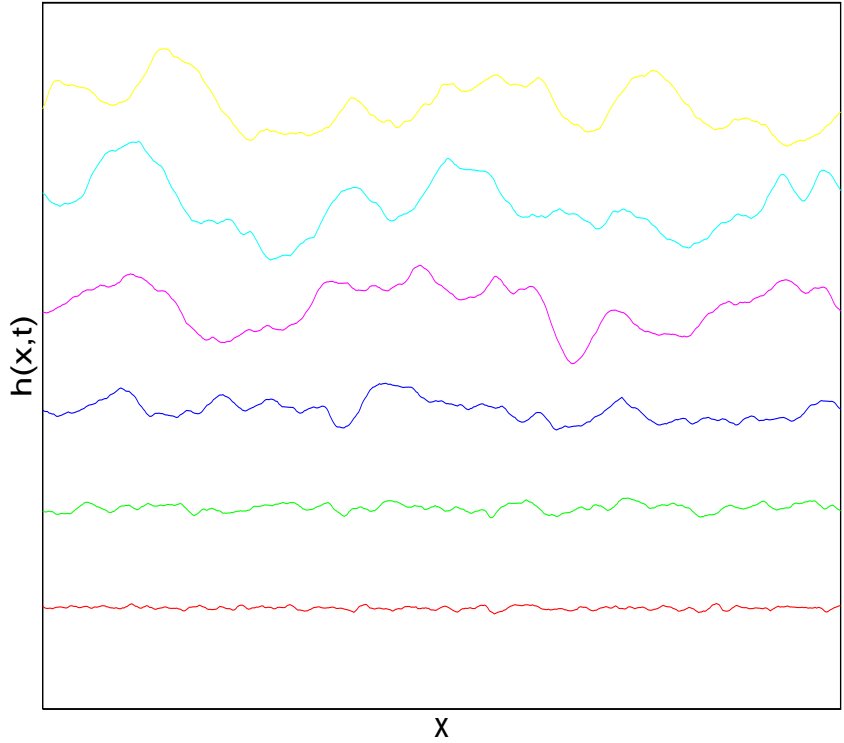}}
\caption{A part of the rough height profile at different times (bottom to top) t\ =\ 10$^3$ - 10$^8$ time
steps . Here $D_h$=$D_\rho = 1$ and $\mu = 1$, $\nu = 0.01$ }
\label{height}
\end{center}
\end{figure}

\n We should note that the above expansion is not well-defined, as the coefficients of the
expansion may themselves be large or divergent. As in \cite{bmb}, we shall still go forward
in the spirit of self-consistency, $i.e.$ subject to numerical verification. The Heaviside
function introduces non-linearities. One way to gain some insight is to carry out a Hartree
type mean-field approximation and replace the non-linearities by their expectation values.
This leads to equations :

\begin{eqnarray}
\fl \frac{\partial h(x,t)}{\partial t}  \eq  -\ D_h\ \frac{d^4 h(x,t)}{dx^4}\ -\ \hat{\nu}\ \frac{d^2 h(x,t)}{dx^2}\ +\ \hat{\mu}\ \rho(x,t)\frac{d^2 h(x,t)}{dx^2}\pls \eta_h(x,t)\nonumber \\    
\fl\phantom{x} \nonumber\\
\fl \frac{\partial \rho(x,t)}{\partial t}  \eq  \phantom{-}\ D_\rho\ \frac{d^2 \rho(x,t)}{dx^2}\ +\ \hat{\nu}\ \frac{d^2 h(x,t)}{dx^2}\ -\ \hat{\mu}\ \rho(x,t)\frac{d^2 h(x,t)}{dx^2} \label{hartree}     
\end{eqnarray}

\n where, $\hat{\mu}= c\mu$ and $\hat{\nu}= (1-c)\nu$. Similar approximations have been studied
by Bouchaud \etal (\cite{bou}). We expect that in some regime our equations (\ref{req}) will
reproduce the mean field results suggested by equations (\ref{hartree}).

\begin{figure}[t]
\begin{center}
\epsfxsize=2.5in \epsfysize=3.in
\rotatebox{270}{\epsfbox{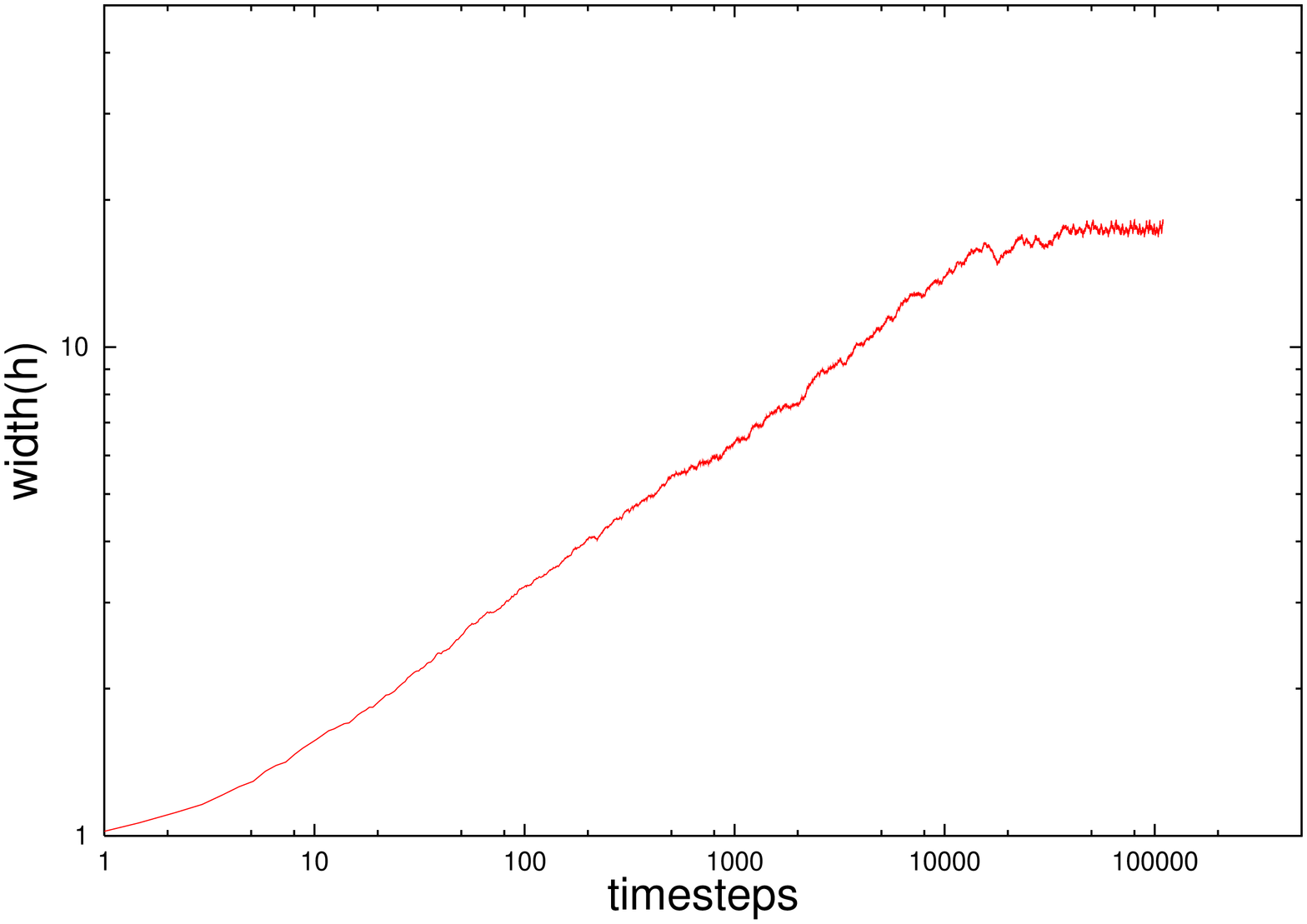}}
\rotatebox{270}{\epsfbox{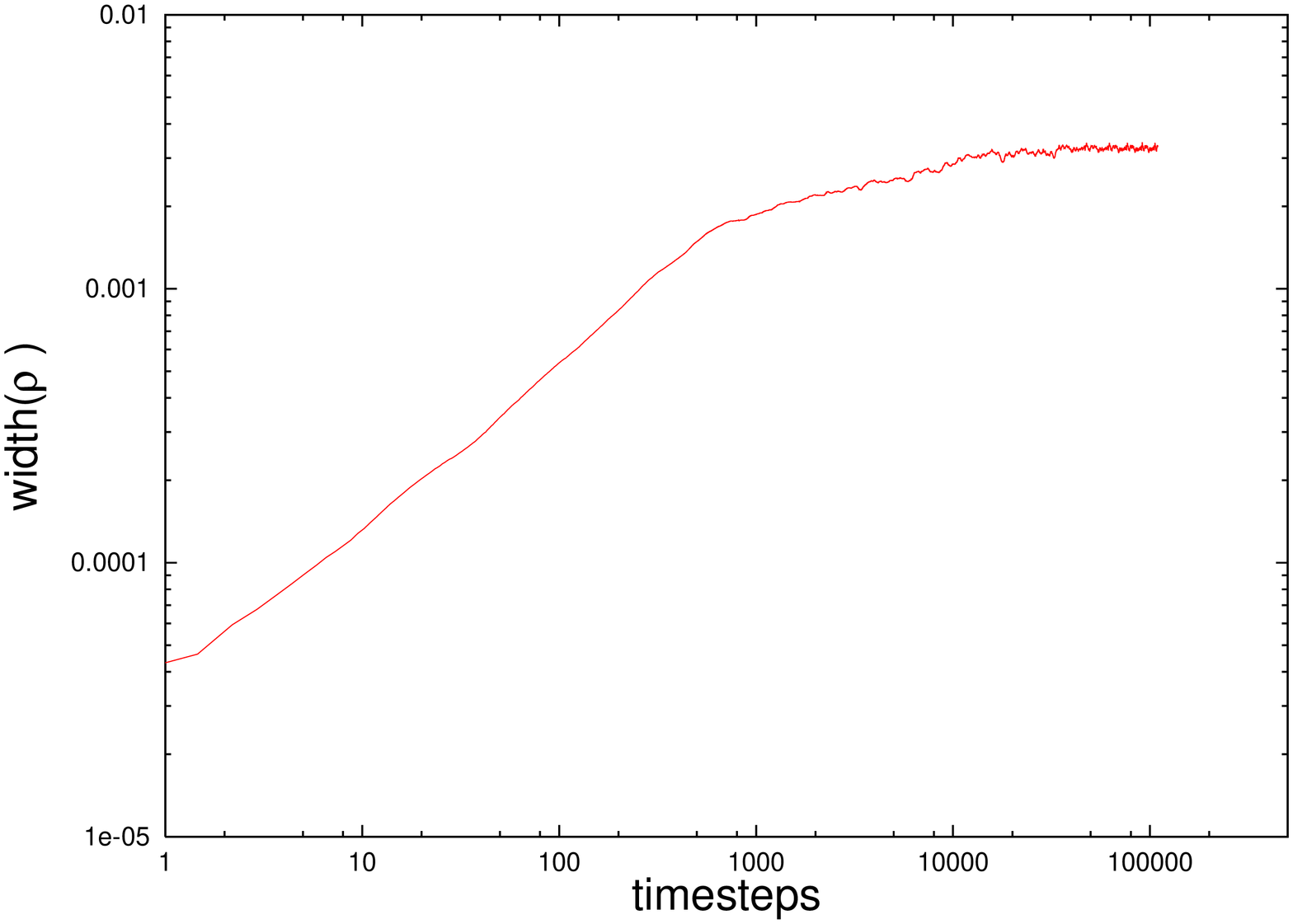}}
\caption{The height(left) and density (right) root mean square deviations as functions of time, showing saturation}
\label{auto}
\end{center}
\end{figure}

\n Figure \ref{height} shows a part of the growing rough height profile for $D_h = D_\rho = 1$, $\mu = 1$ and
$\nu = 0.01$. The heights have been scaled in order to bring out the detailed features for comparison. 
We note that with increasing time short length-scale features slowly die out and mounds and grooves spanning
longer lengths are formed. Figure \ref{auto} shows the variation of the root-mean square height deviation
and density as a function of time. Both quantities show saturation with time. The full dynamical description
then involves {\sl both} the growing and the saturated profiles.

\noindent The double Fourier transform is  defined as :

\[ h(k,\omega) \eq \int dx\ \int dt\ \exp \{-i\ (kx-\omega t)\}\ h(x,t) \]

\noindent The Green functions of the linear part of the equations are :

\begin{eqnarray}
 \left\langle \frac{\delta \tilde{h}(k,\omega)}{\delta\eta(k',\omega')}\right\rangle \eq G_h(k,\omega)\ \delta(k+k')\ \delta(\omega+\omega') \nonumber\\
\phantom{x}\nonumber\\
\left\langle \frac{\delta \tilde{\rho}(k,\omega)}{\delta\eta(k',\omega')}\right\rangle \eq G_\rho(k,\omega)\ \delta(k+k')\ \delta(\omega+\omega') 
\end{eqnarray}

\noindent The correlation functions are given by :

\begin{eqnarray}
S_h(k,\omega) \eq \langle h(k,\omega)\ h(-k,-\omega)\rangle\nonumber\\
S_\rho(k,\omega) \eq \langle \rho(k,\omega)\ \rho(-k,-\omega)\rangle
\end{eqnarray}

Now from equations (\ref{req}),  retaining only the linear terms,
we obtain the two Green functions for $h$ and $\rho$,

\begin{eqnarray}
g_h(k,\omega) \eq \left(-i\ \omega\pls D_h k^4\mns\nu k^2\right)^{-1} \nonumber\\
g_\rho(k,\omega) \eq \left( -i\ \omega\pls D_\rho k^2\right)^{-1} 
\end{eqnarray}

In order to describe the scaling behaviour of the deposition process, we usually define the
following scaling indeces, from the correlation functions :

\begin{eqnarray*}
 S_h(x-x',t-t') \eq \langle h(x,t)h(x',t')\rangle \mns \langle h(x,t)\rangle\langle h(x',t')\rangle \\
 S_\rho(x-x',t-t') \eq \langle \rho(x,t)\rho(x',t')\rangle \mns \langle \rho(x,t)\rangle\langle \rho(x',t')\rangle 
\end{eqnarray*}

\begin{eqnarray*}
 S_h(x,0) \sim \vert x\vert^{2\alpha_h} \quad\quad 
S_\rho(x,0) \sim \vert x\vert^{2\alpha_\rho} \quad\quad\mbox{    as $\vert x\vert \rightarrow \infty$} \\
S_h(0,t) \sim\  \vert t\vert^{2\beta_h}\quad\quad
\ S_\rho(0,t) \sim\ \vert t\vert^{2\beta_h}\quad\quad\mbox{    as $\vert t\vert \rightarrow \infty$}
\end{eqnarray*}

\n In general :

\begin{eqnarray*}
 S_h(x,t) \simeq \ \vert x\vert^{2\alpha_h}\ F_h\left({\vert t\vert}/{\vert x\vert^{z_h}}\right)\\
 S_\rho(x,t) \simeq \ \vert x\vert^{2\alpha_\rho}\ F_\rho\left({\vert t\vert}/{\vert x\vert^{z_\rho}}\right)
\end{eqnarray*}

\n The scaling functions $F_h(\xi)$ and $F_\rho(\xi)$ are assumed to be universal and the indeces $\alpha_h,\alpha_\rho$
and  $z_h$ = $\alpha_h/\beta_h,\ z_\rho\ =\ \alpha_\rho/\beta_\rho$ are the roughness and dynamical exponents. 
Within the strong scaling hypothesis $z_h=z_\rho$ and there exists a single time scale and a
distance scale.

\n In the presence of two time scales, there is only a weak scaling \cite{bmb} hypothesis available to us :

\begin{eqnarray*}
G_h(k,\omega) \eq k^{-z_{h}}\ \Phi_h\left( \frac{\omega}{k^{z_h}}, \frac{\omega}{k^{z_\rho}}\right) \\
G_\rho(k,\omega) \eq k^{-z_{\rho}}\ \Phi_\rho\left( \frac{\omega}{k^{z_h}}, \frac{\omega}{k^{z_\rho}}\right) \\
\end{eqnarray*}

\n Here, $z_h\ \ne\ z_{\rho}$. Absence of strong scaling means that the exponents $\alpha_h$ and $\alpha_\rho$
may become functions of k. 

\subsection{Scaling analysis for the height-height correlations}
\begin{figure}
\centering
\epsfxsize=5in\epsfysize=5in\rotatebox{0}{\epsfbox{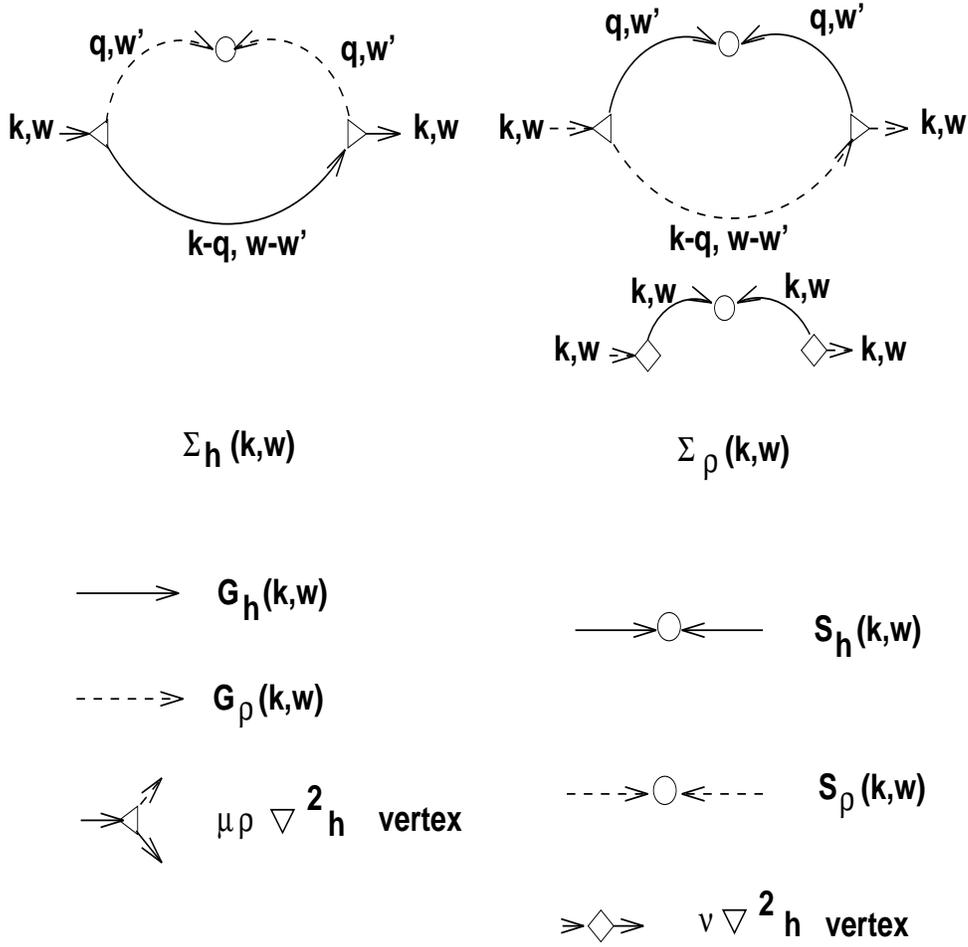}}
\caption{One loop scattering diagrams for the self-energy for the Green functions.
The scattering vertices are shown at the bottom of the figure}
\label{self}
\end{figure}

 Let us look at the one loop diagram for the self-energy \cite{bmb} :

\begin{eqnarray}\fl
\Sigma_h(k,\omega) \eq \mu^2\ \int\frac{dq}{2\pi}\ \int\frac{d\omega'}{2\pi}\ G_h(k-q,\omega-\omega')\ k^2(k-q)^2\ S_\rho(q,\omega')  \nonumber\\
\fl \simeq \mu^2\int\frac{dq}{2\pi}\int\frac{d\omega'}{2\pi}\left[
\frac{1}{-i\ (\omega -\omega')+(k-q)^4-\nu(k-q)^2+\Sigma_h(k-q,\omega-\omega')}\right]\ldots\nonumber\\
\phantom{xxxx}\ldots\frac{k^2(k-q)^2}{q^{1+2\alpha_\rho}} \left[ \frac{\Sigma_\rho(q,\omega')}{\omega'^2 + \Sigma_\rho^2(q,\omega')}\right]
 \nonumber\\
\end{eqnarray}

The integrand over the internal momentum $q$ has a factor $q^{-(1+2\alpha_\rho)}$, which causes an infra-red
divergence in the integral.
As discussed by \cite{bmb}, we take care of the infra-red divergence 
 by introducing a lower cut-off $k_0 \ll 1$. When we talk about ``small values of momenta", we have to
mean $k\sim k_0$. The argument for smaller values of $k$ will have follow a different track.

\begin{figure}[t]
\centering
\epsfxsize=4.5in\epsfysize=4.5in\rotatebox{0}{\epsfbox{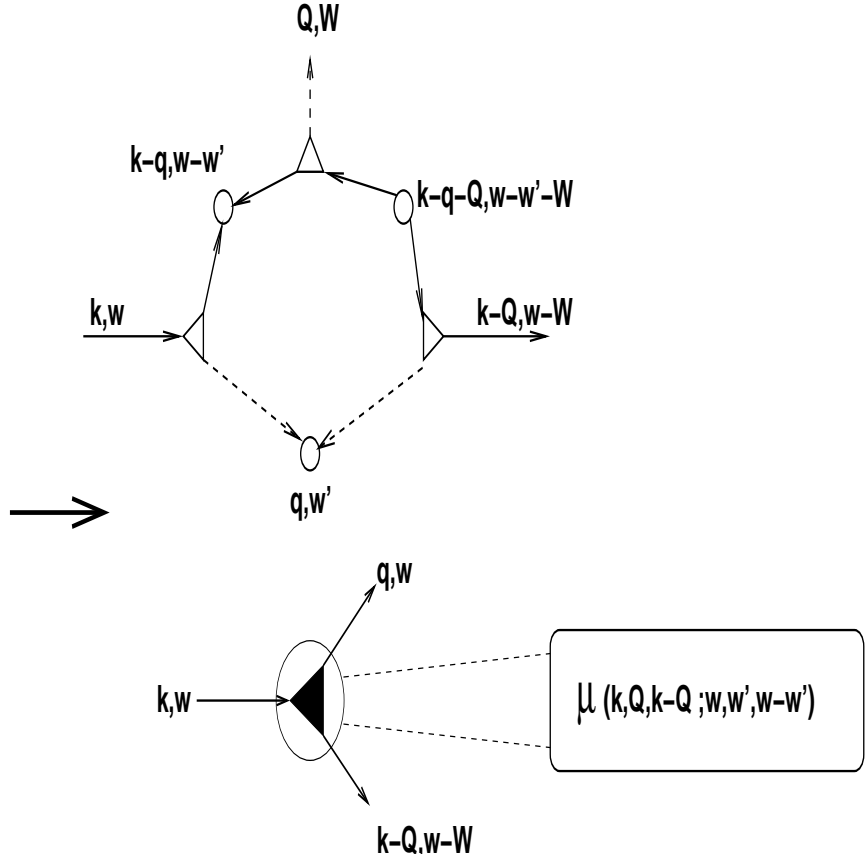}}
\caption{Vertex renormalization}
\label{vertex}
\end{figure}

 Note also that because of the term $q^{-(1+2\alpha_\rho)}$ in the integrand, the main contribution
of the integral comes from small values of $q$. Now, for ``small" internal momenta
$k\gg q\sim k_0\ ;\ \omega'\sim q^{z_h}$, we can replace $G_h(k-q,\omega-\omega') \simeq 
\left( i\ \omega + k^4 -\nu k^2+ \Sigma_h(k,\omega)\right)^{-1}$. We now look at $\omega = 0$, and
noting that $\Sigma(k,0)\sim k^{z_h}$, we note immediately that if $z_h<4$, then the
inverse Green function is dominated by the self-energy :

\[ G^{-1}_h(k,0) \sim \Sigma_h(k,0) \]

\n Substituting this back in the equation for the self-energy :

\[ \Sigma_h(k,0) \simeq \frac{\mu^2k^4}{\Sigma(k,0)} \ A_\rho
 \int_{k_0}^{\infty}\frac{dq}{2\pi}\ \int\frac{d\omega'}{2\pi}\ \frac{1}{q^{1+2\alpha_\rho}} 
\]

This gives :

\[ \Sigma_h^2(k,0) \simeq \frac{\mu^2k_0^{-2\alpha_\rho}}{8\pi\alpha_\rho}\ k^4 \]

\n Since, by definition $\Sigma_h(k,0)\sim k^{z_h}$, it follows immediately that $z_h\ =\ 2$
\vskip 0.1cm

\n Here we have considered only the bare one-loop diagram. In general, the scattering terms
leads to a renormalization of the $\mu$  vertex through the introduction
of  vertex functions $\mu(k,q,k-q)$ for $\omega =0$. The renormalization is
illustrated in figure \ref{vertex}.
 For $q\rightarrow 0$, we may assume $\mu(k,0,k) \sim k^{x_\mu}$ .
Putting these back into the equation for the self-energy, we note that $z_h$ is renormalized
to $z_h + \delta$ where $\delta = x_\mu/2$. The exact numerical values for
$z_h$ may then differ from 2.

\vskip 0.2cm

\begin{figure}[t]
\centering
\epsfxsize=3.0in\epsfysize=2.5in\rotatebox{0}{\epsfbox{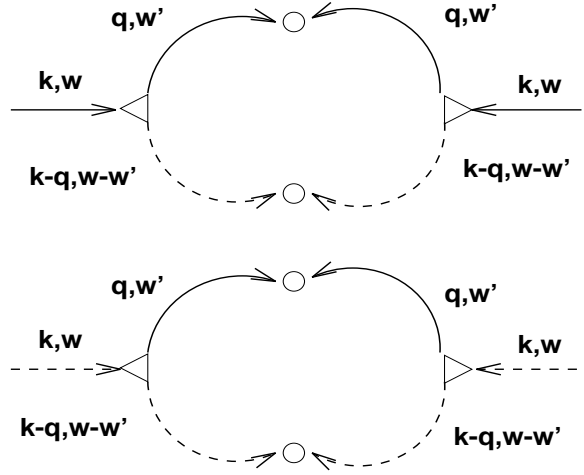}}
\caption{One-loop diagrams for the height-height and density-density correlation functions}
\label{hh}
\end{figure}

The one-loop correction to the height-height correlation function is shown in figure (\ref{hh}). We may immediately
write :

\begin{eqnarray*}
\fl S_h(k,\omega) \eq \frac{1}{\omega^2+\vert\Sigma_h(k,\omega)\vert^2}\ \left[ 1 + \mu^2\ \int \frac{dq}{2\pi}\int
\frac{d\omega'}{2\pi}\ (k-q)^4 \ S_h(k-q,\omega-\omega')\ S_\rho(q,\omega')\right] 
\end{eqnarray*}

\n Let us first examine the  terms on the right-hand side :

\begin{eqnarray*}
\fl\eq  \frac{1}{\omega^2+\vert\Sigma_h(k,\omega)\vert^2}\left[ 1 +\mu^2 \ \int \frac{dq}{2\pi}\int
\frac{d\omega'}{2\pi}\ \frac{(k-q)^4}{(k-q)^{1+2\alpha_h}}\ \left(\frac{1}{q^{1+2\alpha_\rho}}\right)\ \ldots\right.\\ 
\fl \left.\ldots\phantom{xxxxxx}\left(\frac{\Sigma_\rho(q,\omega')}{\omega^2+\vert\Sigma_\rho(q,\omega')\vert^2}\right)
\left(\frac{\Sigma_h(k-q,\omega-\omega')}{(\omega-\omega')^2+\vert\Sigma_h(k-q,\omega-\omega')\vert^2}\right)\right]
\end{eqnarray*}

\vskip 0.4cm
\n Going back to the expression for the self-energy, we see that :

\[ \Sigma_h(k,\omega) \simeq \frac{\Gamma_0^2k^4}{-i\ \omega+\Gamma_0k^2}\]

\n where

\[ \Gamma_0^2 \eq \frac{\mu^2k_0^{-\alpha_\rho}}{8\pi\alpha_\rho} \]

\n Substituting this in the expression for the correlation,

\be S_h(k,\omega) \eq \left(\omega^2+\frac{\Gamma_0^4k^8}{\omega^2+\Gamma_0^2k^4}\right)^{-1}\left(1+\frac{\Gamma_0 k^2}{\omega^2+
\Gamma_0^2k^4}\right)\label{dft}\ee

\n This is our main result from which various limits may be obtained. For instance, 
putting $\omega$ = 0 :

\begin{eqnarray*}
S_h(k,\omega=0) \sim  \frac{1}{\Gamma_0^2k^4}\left( 1+\frac{1}{\Gamma_0k^2}\right)\\
\phantom{xxxxxxxxxxx} \sim k^{-6} \quad\quad\mbox{for small } k 
\end{eqnarray*}

\begin{figure}[t]
\centering
\epsfxsize=3in\epsfysize=4.5in\rotatebox{270}{\epsfbox{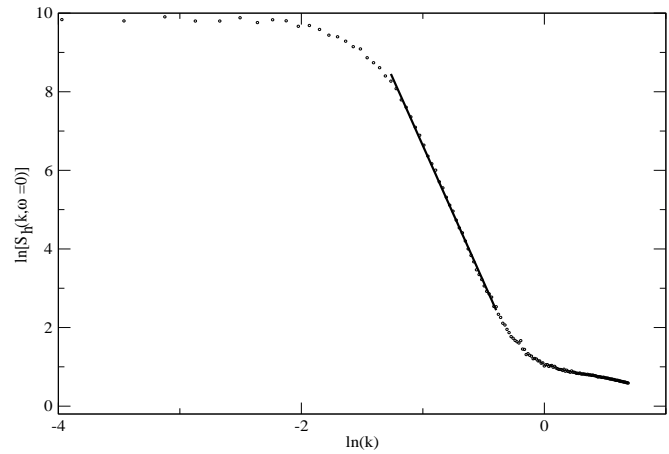}}
\caption{The Log-log plot of the double Fourier transformation $S_h(k,\omega=0)$ $vs$ k.
For small k the slope is $-1-2\alpha_h-z_h$ $\simeq$ --6.75} 
\label{h1}
\end{figure}

\n The numerical results are shown in figure \ref{h1}. The best fit slope was
found to be 6.75. Now, using the expression

\[ S_h(k,\omega=0) \sim k^{-(1+2\alpha_h+z_h)} \]

\n we obtain an equation :

\be 1\pls 2\ \alpha_h\pls z_h \eq 6.75 \label{ind1} \ee

\begin{figure}[b]
\centering
\epsfxsize=3in\epsfysize=4.5in\rotatebox{270}{\epsfbox{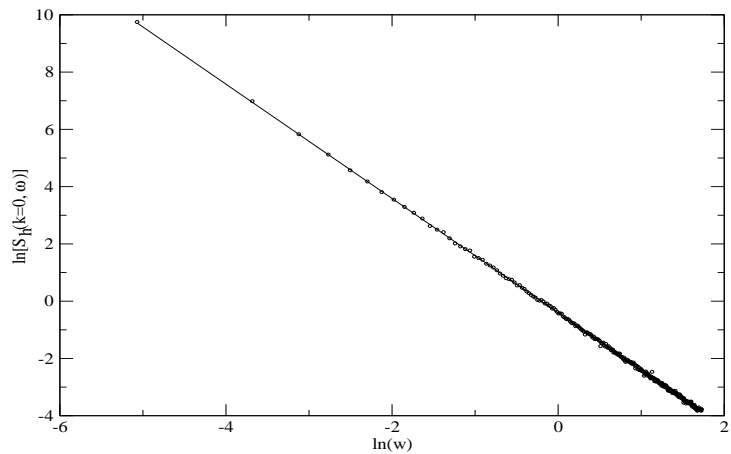}}
\caption{The Log-log plot of the double Fourier transformation $S_h(k=0,\omega)$ $vs$ $\omega$.
For small $\omega$ the slope is $-1-2\beta_h-1/z_h$ $\simeq$ -1.996} 
\label{h2}
\end{figure}

\n Again, putting k=0 and referring to figure \ref{h2} we get :

\[ S_h(0,\omega) \sim \omega^{-2} \]

\n Numerically this has been found to be 1.996.  Using the expression,

\[ S_h(k=0,\omega) \sim \omega^{-(1+2\beta_h+1/z_h)} \]

\n we get another equation,

\be 1\pls 2\ \beta_h\pls 1/z_h \eq 1.996 \label{ind2} \ee

\n Using the equations (\ref{ind1}) and (\ref{ind2}) and $z_h\eq \alpha_h/\beta_h $ we
estimate :

\be \alpha_h \eq 1.2 \quad\quad \beta_h\eq 0.36 \quad \mbox{and}\quad z_h\eq 3.38\ee

\n Note that numerical estimate of $z_h$ is greater than the one-loop estimate of 2.
As discussed earlier, the one-loop estimate is  perturbative and
 may be considerably modified by vertex renormalization,
as well as higher order diagrams.
\begin{figure}[b]
\centering
\epsfxsize=2.5in\epsfysize=4in\rotatebox{270}{\epsfbox{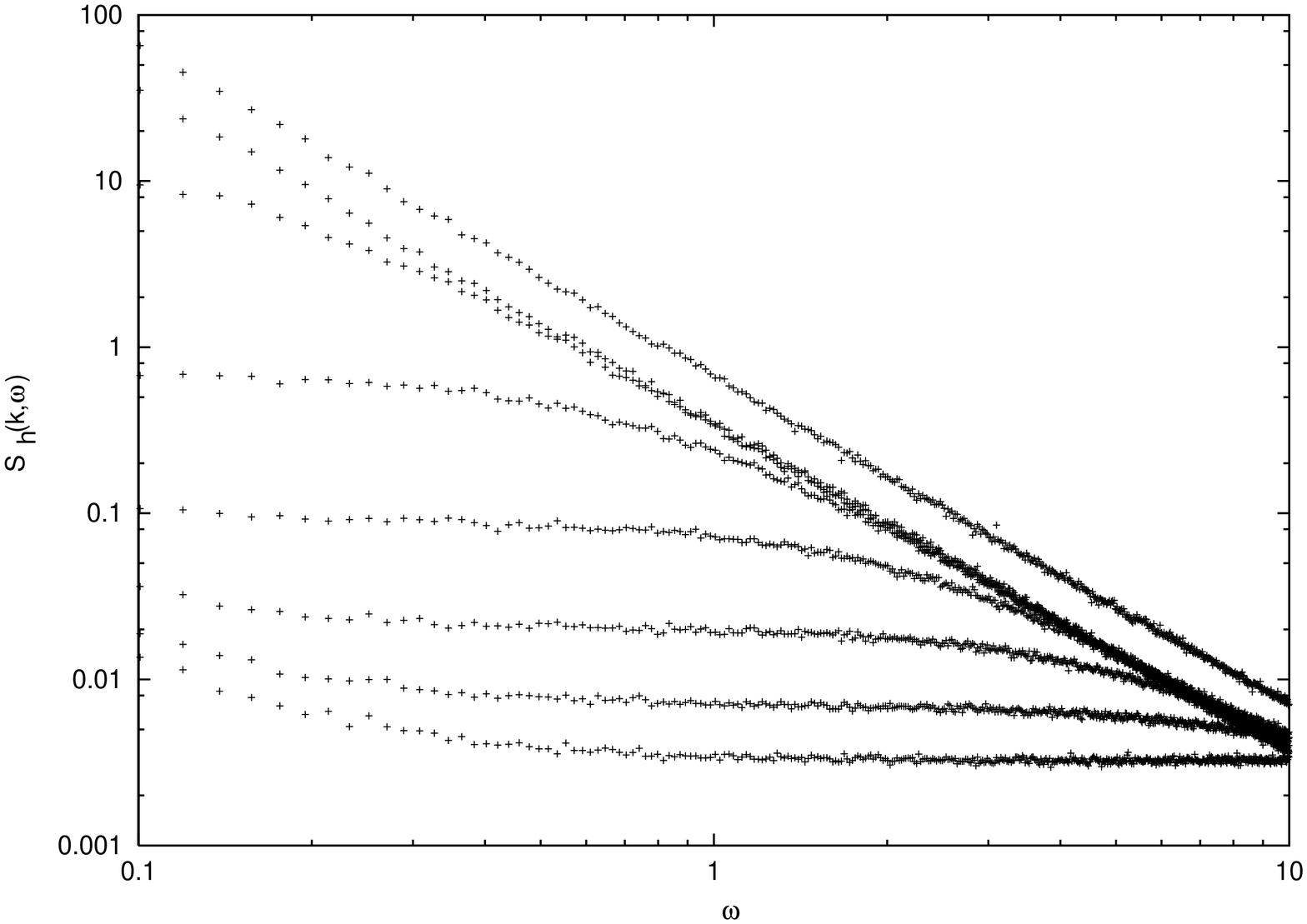}}
\epsfxsize=2.5in\epsfysize=4in\rotatebox{270}{\epsfbox{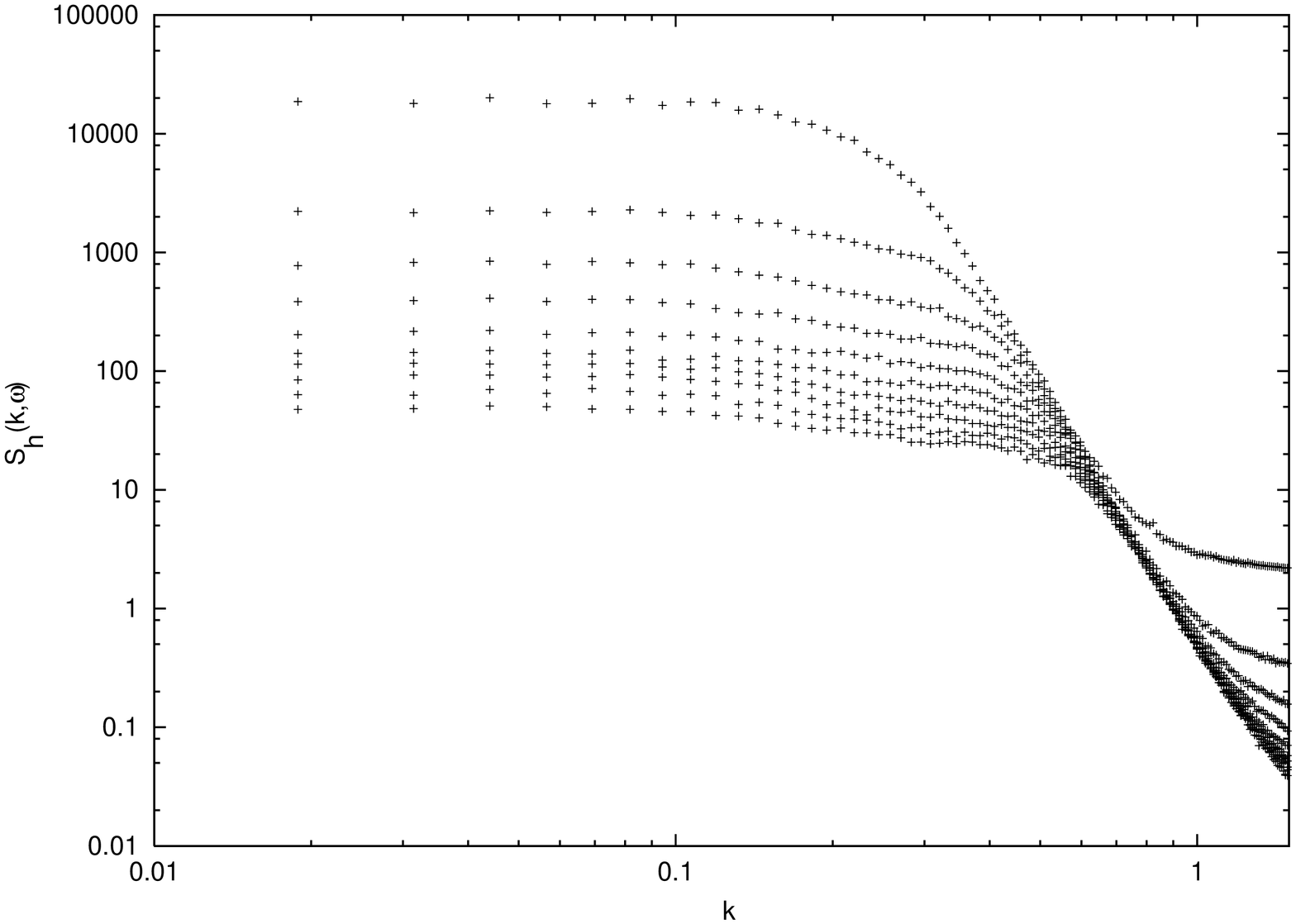}}
\caption{The Log-log plot of the double Fourier transformation $S_h(k,\omega)$ $vs$ (top)
$\omega$ for different values of k and (bottom) $vs$ k for different values of $\omega$ .
}
\label{h3}
\end{figure}

We get further information from the
full double Fourier transform. For example, for sufficiently small k, but $\omega\ \ne\ $0  we get,

\[ S_h(k,\omega) \sim A(\omega) + B(\omega)\ k^2 \]

\n As $k \rightarrow 0$ this leads to a flattening of the curve to a $\omega$-dependent
constant $A(\omega)$. The value of $k$ at which the flattening occurs goes on decreasing
as $\omega$ decreases. 
 Similarly, for $k\ \ne\ $0 , for sufficiently small $\omega$ :

\[ S_h(k,\omega) \sim C(k) + D(k)\ \omega^2 \]

\n As before, the value of $\omega$ at which the flattening occurs decreases as we decrease
k. This crossover behaviour is seen in our numerical results. It is easy to observe that we
would not have gleaned this information from single Fourier transforms, a point emphasized
earlier in the work of Biswas \etal (\cite{bmb}). We may interpret this as a long-time,
smoothening of the growing surface and is a characteristic of the evaporation-accretion process.

\begin{figure}
\centering
\epsfxsize=2.5in\epsfysize=4in\rotatebox{270}{\epsfbox{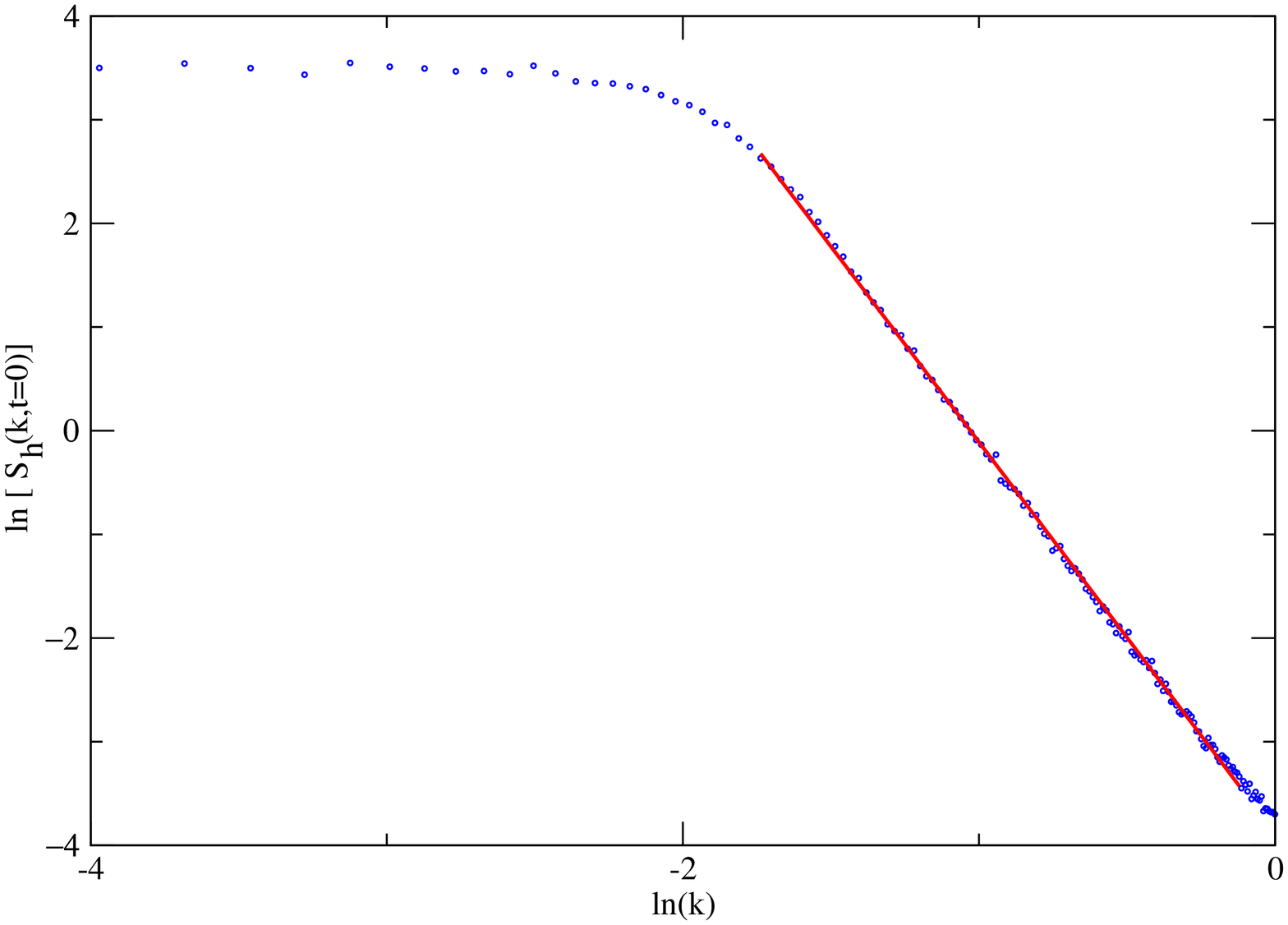}}
\epsfxsize=2.5in\epsfysize=4in\rotatebox{270}{\epsfbox{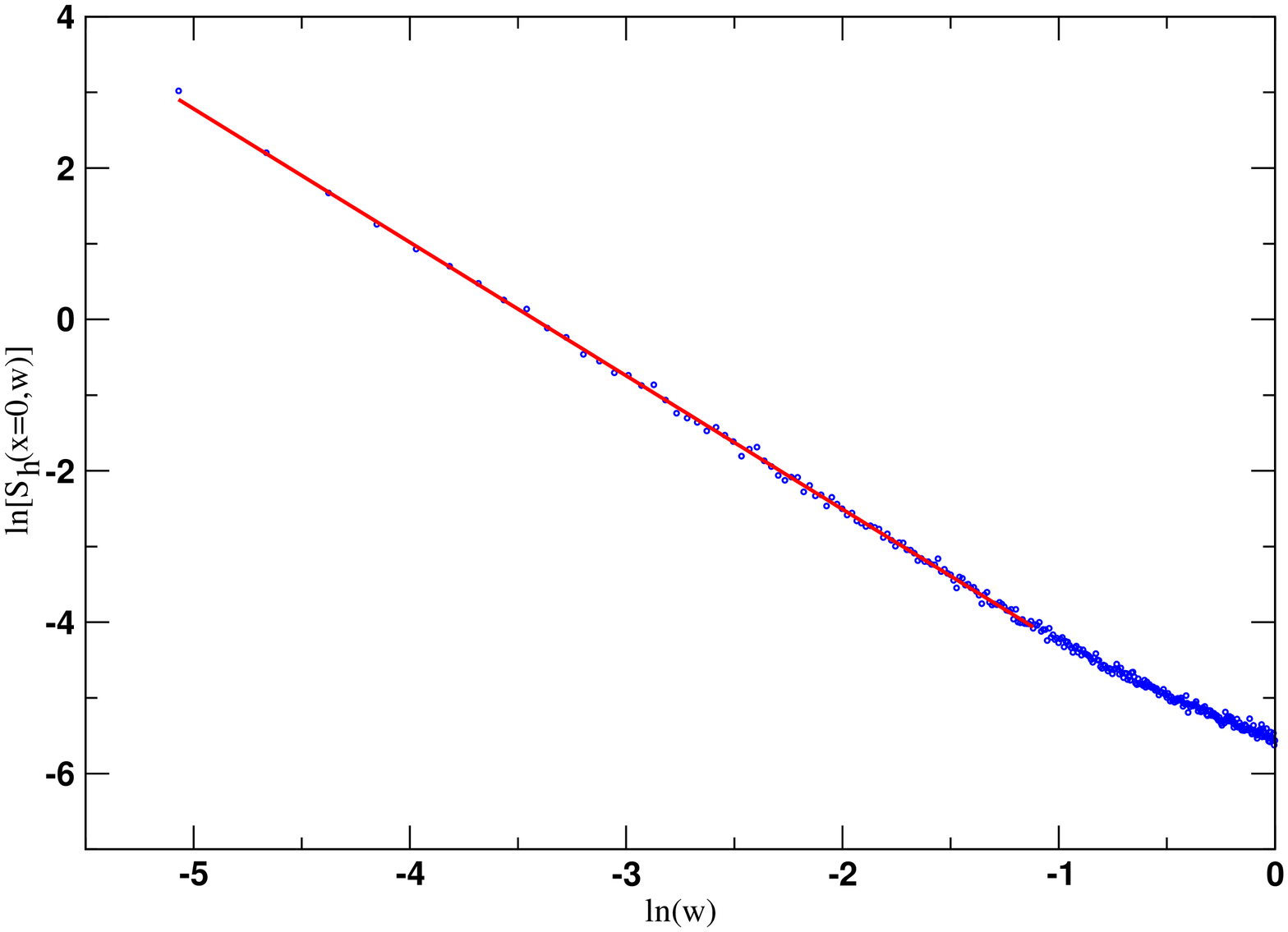}}
\caption{The Log-log plot of the single Fourier transformation (top) $S_h(k,t=0)$ $vs$ $k$, fitted to the slope $-1-2\alpha_h\eq -3.746$ 
 and (bottom) $S_h(x=0,\omega)$ $vs$ $\omega$ fitted to the slope $-1-2\beta_h\eq -1.763$
}
\label{sft1}
\end{figure}

\n In the range of $k$ prior to flattening, we may derive :

\begin{eqnarray*}
S_h(k,t=0) &\eq &\int \frac{d\omega}{2\pi}\ S_h(k,\omega) \\
&\simeq & \frac{1}{2\pi}\int\ \frac{d\omega}{\omega^2+k^{2z_h}} \ +\ \mu^2\ k^{3-2\alpha_h+z_h}\int\
 \frac{d\omega}{(\omega^2+k^{2z_h})^2} \\
&\sim &  A\ k^{-z_h} + B\ k^{3-2\alpha_h-2z_h} 
\end{eqnarray*}

\n Estimates of $\alpha_h$ and $\beta_h$ are 1.37 and 0.38 respectively, consequently
$z_h \eq $ 3.61. Note that
there is no reason why the estimates from single Fourier transforms and the
double Fourier transform should agree exactly. For the single Fourier transforms
we need information only about the saturated surface for $S_h(k,t=0)$ and the
growing surface for $S_h(x=0,\omega)$. For the double Fourier transform, we
need information for both the saturated and growing surfaces for the same
function. In case there are multiple length scales in the problem that information
will be reflected in the double Fourier transform \cite{bmb}.

\n From figure \ref{sft1} in the $k\ll k_0$ region, flattening suggests that :

\[ S_h(k,t=0)\sim const \sim k^0 \]

\n Since  $S_h(k,t=0)\sim k^{-(1+2\alpha_h)}$, the numerical results suggest that in this regime
$\alpha_h = -0.5$ and $z_h = 0$. In this regime, we have $S_h(k,\omega=0) \sim k^{-(1+2\alpha_h+z_h)}$
$\sim$ $k^0$. This crossover to a flattened regime for very small $k$ is also seen in the numerical 
results of figure \ref{h1}.

\subsection{Scaling relations for density-density correlations}

Again referring back to figure (\ref{self}), we can write an expression for the  self-energy for the density Green function :

\begin{eqnarray*}
\fl \Sigma_\rho(k,\omega)\eq \mu^2\int\frac{dq}{2\pi}\int\frac{d\omega'}{2\pi}\ G_\rho(k-q,\omega -\omega')\ S_h(q,\omega')k^2(k-q)^2 +
 \nu^2k^4 S_h(k,\omega)\\
\fl\eq \mu^2\int\frac{dq}{2\pi}\int\frac{d\omega'}{2\pi}\ 
\left(\frac{1}{-i\ (\omega-\omega')+\Sigma_\rho(k-q,\omega-\omega')} \right)\ 
\frac{1}{q^{1+\alpha_h}}\ldots\\
\ldots \left(\frac{1}{\omega'^2+\vert\Sigma_h(q,\omega')\vert^2} k^2(k-q)^2\right)
 +   \nu^2k^4 S_h(k,\omega)
\end{eqnarray*}

\begin{figure}
\centering
\epsfxsize=3in\epsfysize=4in\rotatebox{270}{\epsfbox{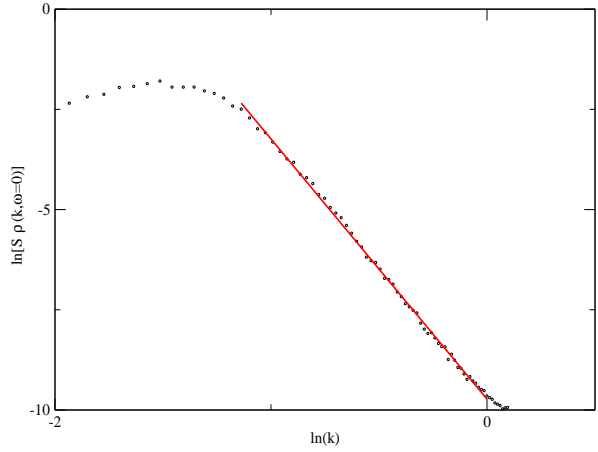}}
\caption{Log-log plot of the double Fourier transform $S_\rho(k,\omega=0)$ $vs$ k.
The best fit slope is : $-1-2\alpha_\rho-z_\rho$\eq --6.49.}
\label{rho1}
\end{figure}

\n We may now carry out the integral over $\omega'$ and use the fact that the integrand over $q$ has a
factor $q^{-(1+\alpha_h)}$ which ensures that only small $q$ values contribute to the integral. This,
combined with a lower cut-off for $q$, $k_0$ allows us to evaluate the $q$ integral approximately. The
procedure is almost exactly like the case of height-height correlations :

\begin{eqnarray*}
 \Sigma_\rho(k,\omega) \simeq  \frac{\Gamma_1^2 k^4}{-i\ \omega + \Gamma_1k^2}\\
\end{eqnarray*}

\n For small $k$, we know that $\Sigma_\rho(k,\omega=0)\ \sim\ k^{z_\rho}$, and
from above equation :

\[ \Sigma_\rho(k,\omega=0) \simeq A\ k^2 \pls B\ k^{3-2\alpha_h-z_h} \]

\n With $\alpha_h = 1.5$ and $z_h = 2$, we obtain $z_\rho = 2$

	For the density-density correlation function we get (again from figure (\ref{hh})) :

\begin{eqnarray*}
\fl S_\rho(k,\omega)\eq \frac{1}{\omega^2+\vert\Sigma_\rho(k,\omega)\vert^2}\left[ 1+\mu^2
\int\frac{dq}{2\pi}\int \frac{d\omega'}{2\pi}\ S_h(k-q,\omega-\omega')S_{\rho}(q,\omega')\vert k-q\vert^4\ 
\right]\ldots\\
\phantom{xxxxxxxxxxx}+\ \frac{\nu^2k^4}{\omega^2+k^{2z_\rho}}\ S_h(k,\omega) 
\end{eqnarray*}

\begin{figure}
\centering
\epsfxsize=3in\epsfysize=4in\rotatebox{270}{\epsfbox{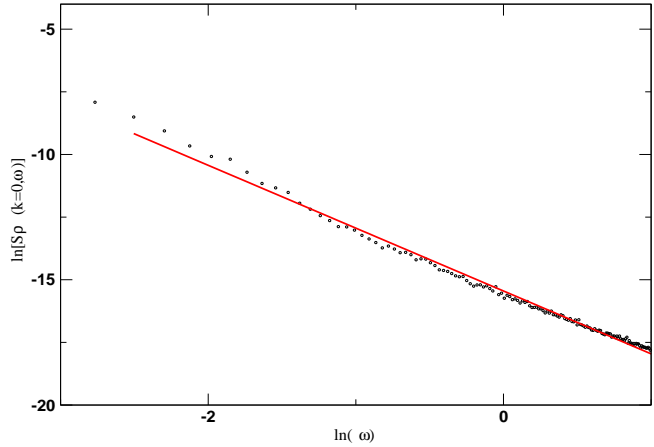}}
\caption{Log-log plot of the double Fourier transform $S_\rho(k=0,\omega)$ $vs$ $\omega$.
The best fit slope is : $-1-2\beta_\rho-1/z_\rho$\eq -2.48.}
\label{rho2}
\end{figure}

\n The first integral over $q$ has a integrand with a factor $q^{-(1+\alpha_\rho)}$, which makes sure that the
main contribution to the integral comes only from small values of $q$. So in the integrand, we can replace 
$k-q$ by $q$ and carry out the $\omega'$ integral, to obtain :

\be S_\rho(k,\omega) \simeq A\ \left(\omega^2+\frac{\Gamma_1^4k^8}{\omega^2+\Gamma_1^2k^4}\right)^{-1}\
\left(1 + \frac{\Gamma_0k^2}{\omega^2+k^4}\right) \ee

\n Given $z_h=2$, $z_\rho = 2$ and $\alpha_h = 1.5$.

\[ S_\rho(k,\omega=0)\sim k^{-6}\]

\n From figure \ref{rho1}, numerically we get an index of $6.49$. 

\vskip 0.2cm
\n In the cross-over regime when $z_h = 0$ , $\alpha_h = -0.5$ and
$z_\rho = 2$:

\[S_\rho(k,\omega=0) \sim  \frac{k^{(3-2\alpha_h+z_h)}}{k^{(2z_h+2z_\rho)}}\sim k^0 \]

\n This is seen numerically  as a flattening in the very low $k$ regime in figure \ref{rho1}.

Again, integrating over the variable $\omega$ we get :

\begin{figure}
\centering
\epsfxsize=3in\epsfysize=4in\rotatebox{270}{\epsfbox{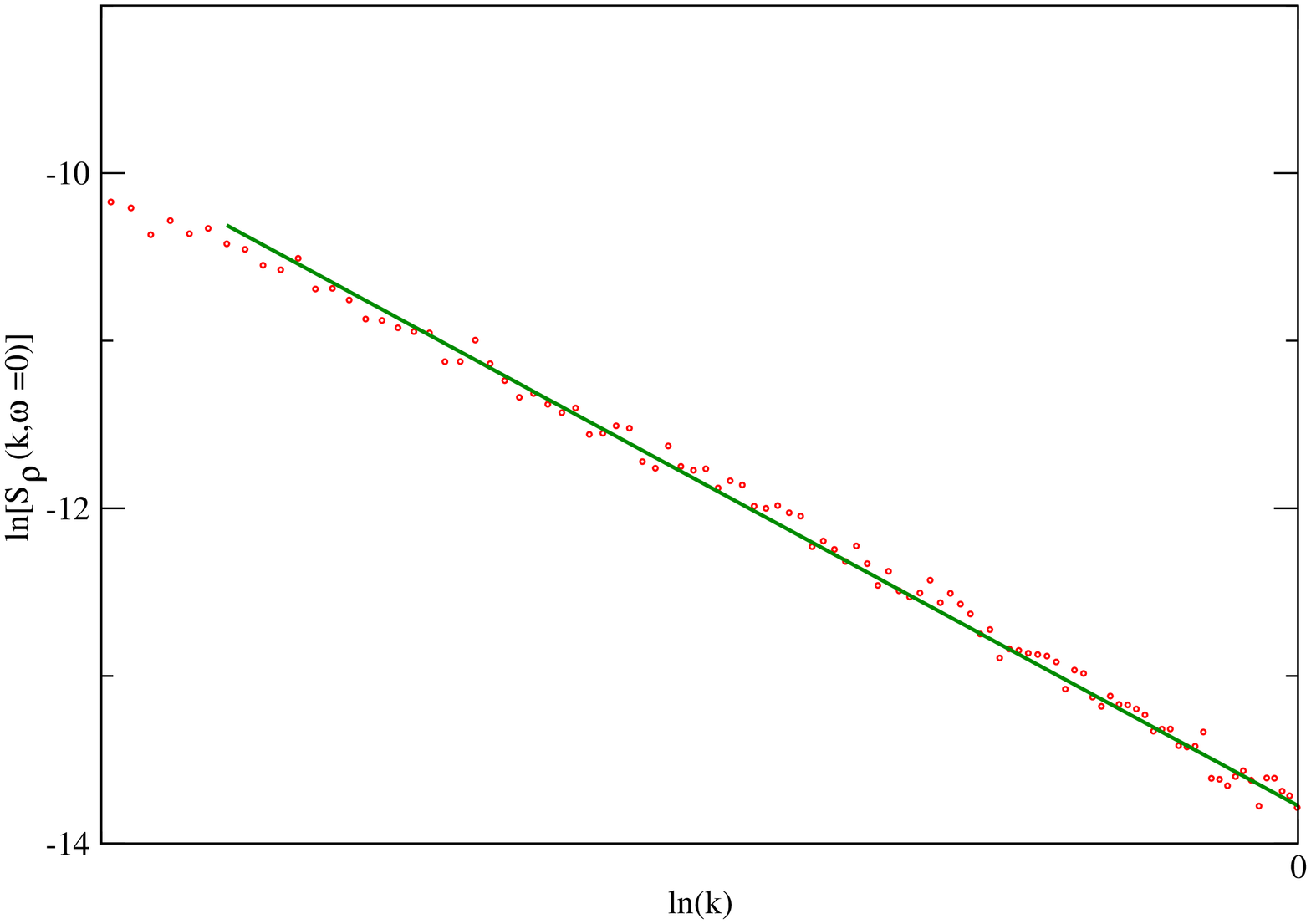}}
\epsfxsize=3in\epsfysize=4in\rotatebox{270}{\epsfbox{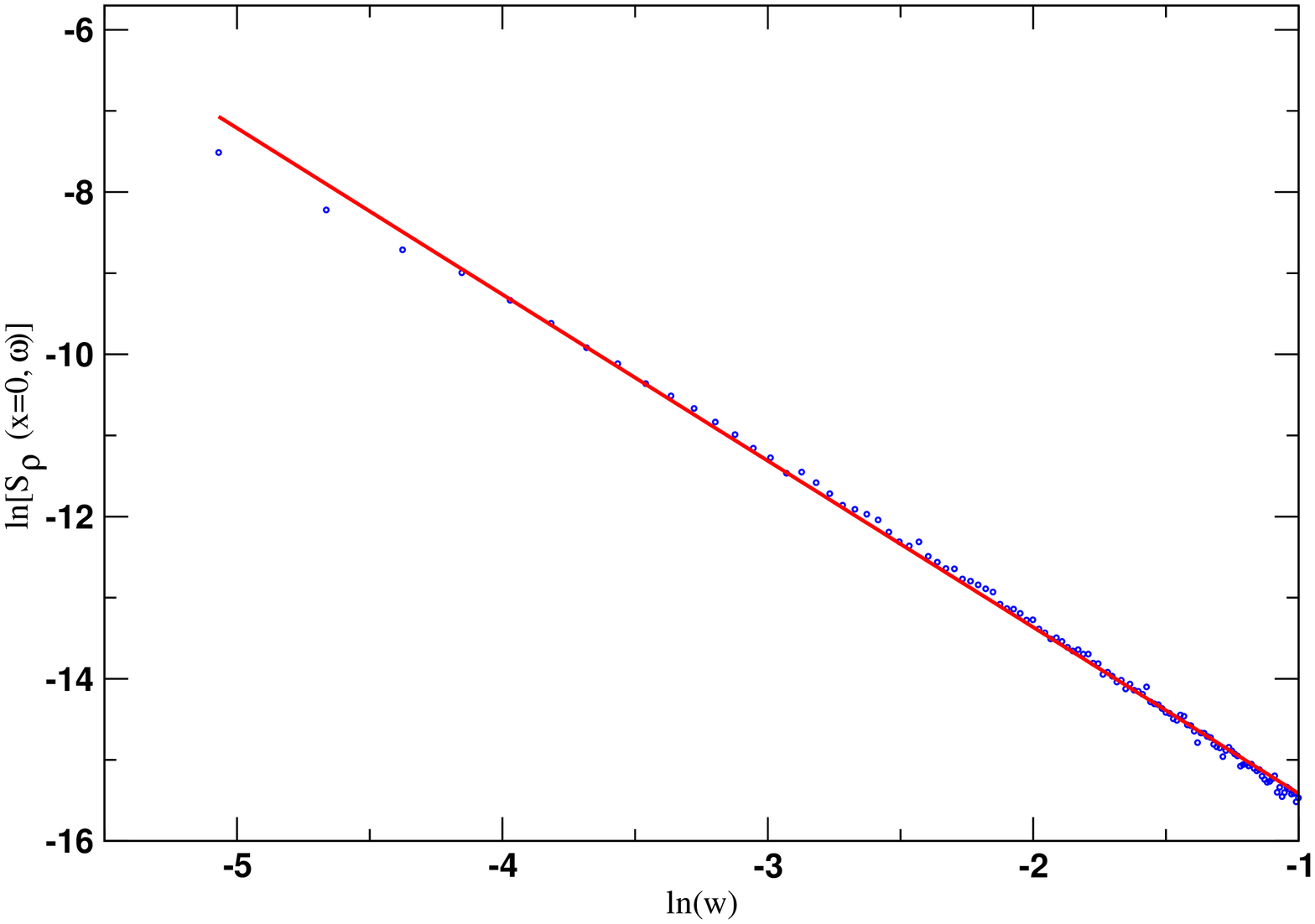}}
\caption{Log-log plot of the single Fourier transform (top) $S_\rho(k,t=0)$ $vs$ k. The
best fit slope is -3.868 (bottom) $S_\rho(x=0,\omega)$ $vs$ $\omega$. The best fit
slope is -2.050}
\label{rho3}
\end{figure}

\begin{eqnarray*}
S_{\rho}(k,t=0) &\eq & \int \frac{d\omega}{2\pi}\ S(k,\omega) \\
& \eq & (\Gamma_0^2/2)\ \frac{1}{k^{z_\rho}(k^{z_h}+k^{z_\rho})} \\
& \sim & k^{-4} \\
\end{eqnarray*}

\n Numerically we find the index to be 3.868.

From the expression for $S_\rho(k,\omega)$ we note that :

\[ S_\rho(k=0,\omega) \sim w^{-2} \]

\n Numerically we obtain an index of 2.488. We may also carry out the integral over $k$ to get :

\begin{eqnarray*}
S_\rho(x=0,\omega) &\eq & \int\frac{k}{2\pi}\ S_\rho(k,\omega) \\
&\simeq & A + B\omega^{-2} 
\end{eqnarray*}

The numerical prediction for the index is 2.05. The accompanying table summarizes our results.

\begin{table}[t]
\centering
\begin{tabular}{||c|c|c||c|c|c||}\hline\hline
Expression & Analytical & Numerical &Expression & Analytical & Numerical \\
           & Index       & Index &          & Index      & Index  \\ 
           & (Single Loop) &     &          & (Single Loop) &      \\ \hline
\phantom{x} & & & & & \\
$\Sigma_h(k,\omega=0)\ vs\ k$ & 2.00   &  3.74 & $\Sigma_\rho(k,\omega=0)\ vs \ k$ & 2.00 &  2.87 \\
$S_h(k,t=0)\ vs\ k$  &  4.00 & 3.75 & $S_\rho(k,t=0)\ vs\ k$ & 4.00 & 3.89 \\
$S_h(k=0,\omega)\ vs\ \omega$ & 2.00 & 1.996 & $S_\rho(k=0,\omega)\ vs\ \omega$ & 2.00 & 2.48 \\
$S_h(k,\omega=0)\ vs\ k$ & 6.00 & 6.65 & $S_\rho(k,\omega=0)\ vs\ k$ & 6.00 & 6.49 \\
\phantom{x} & & & & & \\ \hline\hline
\end{tabular}
\end{table}

\section{Conclusion}

We have studied the scaling behaviour of a set of coupled continuum equations describing
surface growth in the presence of evaporation-accretion both numerically and within a
one-loop perturbative approach. We notice a crossover from  a roughening regime to 
smoothening at very large time, large length scale regimes. The one-loop estimates
gives us an insight into our numerical results. We also notice that in order to study
problems with more than one length-time scales, i.e. if we have weak scaling, it is
essential to study the double Fourier transforms which probe both the growing and the
saturated profiles together. This is in confirmation of the ideas set forward earlier by
Biswas \etal (\cite{bmb}).

\ack
We would like to thank Dr. Anita Mehta for introducing us to the idea of coupled
continuum equations and many discussions about the model we
study in this communication.

\end{document}